\documentclass[12pt,preprint]{aastex}
\usepackage{color}
\usepackage{graphicx}

\slugcomment{Accepted to ApJ}

\shorttitle{Chemical abundances of Kepler-138 and Kepler-186}
\shortauthors{Souto et al.}

\begin{document}

\title{Chemical Abundances of M-dwarfs from the APOGEE Survey. I. The Exoplanet Hosting Stars Kepler-138 and Kepler-186}

\author{D. Souto\altaffilmark{1}}
\author{K. Cunha\altaffilmark{1,2}}

\author{D. A. Garc\'ia-Hern\'andez\altaffilmark{3,4}}

\author{O. Zamora\altaffilmark{3,4}}
\author{C. Allende Prieto\altaffilmark{3,4}}
\author{V. V. Smith\altaffilmark{5}}

\author{S. Mahadevan\altaffilmark{6,7}}
\author{C. Blake\altaffilmark{8,9}}

\author{J. A. Johnson\altaffilmark{10}}
\author{H. J{\"o}nsson\altaffilmark{3,4}}
\author{M. Pinsonneault\altaffilmark{10}}

\author{J. Holtzman\altaffilmark{11}}
\author{S. R. Majewski\altaffilmark{12}}
	
\author{M. Shetrone\altaffilmark{13}}
\author{J. Teske\altaffilmark{14}}

\author{D. Nidever\altaffilmark{15,16,17}}
\author{R. Schiavon\altaffilmark{18}}
\author{J. Sobeck\altaffilmark{12}}

\author{A. E. Garc\'ia P\'erez\altaffilmark{3,4}}
\author{Y. G\'omez Maqueo Chew\altaffilmark{19}}
\author{K. Stassun\altaffilmark{20}}

\altaffiltext{1}{Observat\'orio Nacional, Rua General Jos\'e Cristino, 77, 20921-400 S\~ao Crist\'ov\~ao, Rio de Janeiro, RJ, Brazil}
\altaffiltext{2}{University of Arizona, Tucson, AZ 85719, USA}
\altaffiltext{3}{Instituto de Astrof\'isica de Canarias (IAC), V\'ia Lactea S/N, E-38205, La Laguna, Tenerife, Spain}
\altaffiltext{4}{Departamento de Astrof\'isica, Universidad de La Laguna (ULL), E-38206, La Laguna, Tenerife, Spain}

\altaffiltext{5}{National Optical Astronomy Observatory, 950 North Cherry Avenue, Tucson, AZ 85719, USA}
\altaffiltext{6}{Department of Astronomy and Astrophysics, The Pennsylvania State University}
\altaffiltext{7}{Center for Exoplanets \& Habitable Worlds, The Pennsylvania State University}
\altaffiltext{8}{Department of Physics \& Astronomy, University of Pennsylvania, 209 South 33rd Street,
Philadelphia, PA, USA, 19104}
\altaffiltext{9}{NASA Roman Technology Fellow}

\altaffiltext{10}{Department of Astronomy, The Ohio State University, Columbus, OH 43210, USA}

\altaffiltext{11}{New Mexico State University, Las Cruces, NM 88003, USA} 

\altaffiltext{12}{Department of Astronomy, University of Virginia, Charlottesville, VA 22904-4325, USA}
\altaffiltext{13}{University of Texas at Austin, McDonald Observatory, USA}
\altaffiltext{14}{Department of Terrestrial Magnetism, Carnegie Institution for Science, Washington, DC 20015}

\altaffiltext{15}{Department of Astronomy, University of Michigan, Ann Arbor, MI, 48104, USA}
\altaffiltext{16}{Large Synoptic Survey Telescope, 950 North Cherry Ave, Tucson, AZ 85719}
\altaffiltext{17}{Steward Observatory, 933 North Cherry Ave, Tucson, AZ 85719}

\altaffiltext{18}{Astrophysics Research Institute, Liverpool John Moores University, 146 Brownlow Hill, Liverpool, L3 5RF, United Kingdom}
\altaffiltext{19}{Instituto de Astronom\'ia, Universidad Nacional Aut\'onoma de M\'exico, Ciudad Universitaria, 04510, Ciudad de M\'exico, M\'exico}
\altaffiltext{20}{Department of Physics and Astronomy, Vanderbilt University, Nashville, TN 37235, USA}

\begin{abstract}
We report the first detailed chemical abundance analysis of the exoplanet-hosting M-dwarf stars Kepler-138 and Kepler-186 from the analysis of high-resolution ($R$ $\sim$ 22,500) $H$-band spectra from the SDSS IV - APOGEE survey. Chemical abundances of thirteen elements - C, O, Na, Mg, Al, Si, K, Ca, Ti, V, Cr, Mn, and Fe - are extracted from the APOGEE spectra of these early M-dwarfs via spectrum syntheses computed with an improved line list that takes into account H$_{2}$O and FeH lines. This paper demonstrates that APOGEE spectra can be analyzed to determine detailed chemical compositions of M-dwarfs. Both exoplanet-hosting M-dwarfs display modest sub-solar metallicities: [Fe/H]$_{Kepler-138}$ = -0.09 $\pm$ 0.09 dex and [Fe/H]$_{Kepler-186}$ = -0.08 $\pm$ 0.10 dex. The measured metallicities resulting from this high-resolution analysis are found to be higher by $\sim$0.1-0.2 dex than previous estimates from lower-resolution spectra. The C/O ratios obtained for the two planet-hosting stars are near-solar, with values of 0.55 $\pm$ 0.10 for Kepler-138 and 0.52 $\pm$ 0.12 for Kepler-186. Kepler-186 exhibits a marginally enhanced [Si/Fe] ratio.

\end{abstract}


\section{Introduction}

The Kepler mission (Koch et al. 2010; Batalha et al. 2013) has discovered, to date, more than 2,300 transiting exoplanets around stars of various stellar types. The majority of exoplanets discovered by Kepler have radii between 2 R$_{\oplus}$ $<$ R $<$ 6 R$_{\oplus}$ and most of these are found in close orbits around F-G-K stars, while some 300 have radii smaller than 1.2 R$_{\oplus}$$\footnote{Data from NASA Exoplanet Archive: http://exoplanetarchive.ipac.caltech.edu}$. Given that the detection of exoplanets from their transits is directly related to the size of the planet relative to that of its host star, the cooler and also smaller, low-mass main-sequence stars (M-dwarfs) allow for the detection of smaller planets when compared to solar-type stars.  For example, the smallest exoplanets for which the radius can be measured are hosted by M-dwarfs, with Kepler-138 and Kepler-186 being noteworthy examples. Kepler-138 has two confirmed planets, and is known to host one of the smallest planets discovered to date (Kepler-138b), having been recognized as a Mars-sized exoplanet (Jontof-Hutter et al. 2015), Kepler-186 has five planets and hosts the first Earth-sized exoplanet (Kepler-186f) found in a star's habitable zone, as reported in Quintana et al. (2014). 

A precise, quantitative spectroscopic analysis of these two M-dwarfs, which host particularly interesting exoplanets,  is important for the characterization of the host stars (determination of the effective temperature, $T_{\rm eff}$ , surface gravity, log $g$, metallicity and detailed chemical abundances). The stellar parameters $T_{\rm eff}$ and log $g$ can be used to constrain the stellar radius, which is needed to determine the physical size of an exoplanet, as the transit depth reveals primarily the ratio of exoplanet radius to host-star radius. Knowing the detailed chemistry of the host-star is also important, as this is thought to play a key role in conditions in the protoplanetary disk and subsequent planetary formation.  For example, certain abundance ratios, such as Mg and Si relative to O, can affect core to mantle mass ratios in rocky exoplanets, while C/O ratios control ice chemistry in protoplanetary disks (e.g., Bond et al. 2010; Delgado-Mena et al. 2010; Teske et al. 2014; Thiabaud et al. (2015); Santos et al. (2015); Dorn et al. (2016); Unterborn \& Panero 2016; Brewer \& Fischer 2016). 

Previous detailed spectroscopic analyses of exoplanet host-stars have largely focused on the warmer F-G-K hosts and these analyses have reached impressive levels of precision, with typical precision in $T_{\rm eff}$ of $\sim$ 10-50 K, in log $g$ of $\sim$ 0.1 dex, or in values of elemental abundances ([X/H]) of 0.02 - 0.05 dex in the most precise analyses (e.g., Mel{\'e}ndez et al. 2009; Ghezzi et al. 2010; Adibekyan et al. 2012; Nissen 2015; Schuler et al. 2015).  Such results have yet to appear for exoplanet-hosting M-dwarfs, due primarily to the difficulty of analyzing M-dwarf spectra in the optical region ($\lambda$ $<$ 10,000 \AA{}): dwarfs with $T_{\rm eff}$ $<$ 4000 K exhibit numerous molecular bands, primarily from TiO, that increase dramatically in strength with decreasing $T_{\rm eff}$. However, attempts have been made to derive Fe, Ti and Ca abundances from optical high-resolution spectra (Woolf \& Wallerstein 2005, Bean et al. 2006, Chavez \& Lambert 2009 and Neves et al. 2014). To produce detailed elemental abundance results for M-dwarfs that rival those for F-G-K dwarfs, it is useful to shift the analysis to the near-infrared (near-IR) part of the spectrum ($\lambda$ $\sim$ 1.1 - 2.5 $\mu$m), where both the strength and density of molecular absorption features drops relative to optical wavelengths, especially for the warmer M-dwarfs. Recent high-resolution analyses in the $J$- and $K$- bands have derived metallicities ({\"O}nehag et al. 2012, Lindgren et al. 2016) and C/O ratios (Tsuji et al. 2014, 2015, 2016) in M-dwarfs. 

In this paper, we present the first detailed near-IR chemical abundance analysis of M-dwarfs observed by the SDSS IV - Apache Point Observatory Galactic Evolution Experiment (APOGEE, Majewski et al. 2015). The two targets are the planet hosting M-dwarfs Kepler-138 and Kepler-186. APOGEE is a survey dedicated to studying Galactic evolution from observations of red-giants; however, APOGEE has also observed a number of M-dwarfs (around 2,000) under ancillary projects, or as additional survey targets for filling in some fields. This paper demonstrates that APOGEE spectra can be analyzed to determine detailed chemical compositions of M-dwarfs in general, and of planet-hosting M-dwarfs in particular. Section 2 describes the observed spectra from APOGEE, while in Section 3 the determination of the atmospheric parameters and the abundance analysis for the targets is outlined and presented. Results are discussed in Section 4 and possible connections between certain abundance ratios and exoplanet properties are explored.

\section{APOGEE Spectra of the Exoplanet-Hosting M-dwarfs Kepler-138 and Kepler-186}

The APOGEE spectrograph operating at the Apache Point Observatory (APO) on the SDSS 2.5-meter telescope (Gunn et al. 2006). The instrument is a cryogenic 300-fiber spectrograph covering the wavelength range from  $\lambda$ $\sim$ 1.50 $\mu$m - 1.70 $\mu$m at high resolution ($R$ = $\lambda$/$\Delta$$\lambda$ $\sim$ 22,500; Wilson et al. 2010). The raw data are reduced to wavelength and flux calibrated spectra via pipeline processing (Nidever et al. 2015).

Kepler-138 and Kepler-186 were observed by APOGEE under an ancillary project devoted to the study of M-dwarf radial velocities (Deshpande et al. 2013). APOGEE obtained observations in two visits of three hours each, resulting in combined spectra with final signal-to-noise ratios per pixel of $\approx$ 333 and $\approx$ 96 for Kepler-138 and Kepler-186, respectively. This is the first study to explore in detail high-resolution spectra of M-dwarfs in the APOGEE region. The high S/N spectrum of Kepler-138 (M0.5V) is valuable in identifying atomic and molecular lines that are useful for detailed abundance analyses of M-dwarfs (Figure 1; Section 3.2). These features will then be used to define spectral windows for the automated abundance analyses via the ASPCAP pipeline (APOGEE Stellar Parameter and Chemical Abundances Pipeline; Garc{\'i}a P{\'e}rez et al. 2016). Currently, ASPCAP is designed primarily to analyze the spectra of red-giants, which are the main targets for the APOGEE survey; a future goal will be to have ASPCAP produce reliable results for dwarfs in general, and M-dwarfs in particular. We note that APOGEE data release 12 (DR12; Alam et al. 2015) and earlier do not present stellar parameters and abundances for M-dwarfs (as the DR12 APOGEE line list did not include H$_{2}$O and FeH transitions). The most recent data release 13, (DR13; Albareti et al. 2016) has results for M-dwarfs, however, these are based on a line list that includes H$_{2}$O but not yet FeH transitions, which are important in the spectra of M-dwarfs in this wavelength region (see Section 3.2). A future application of this work will be to incorporate FeH lines into the APOGEE line list.

\section{Atmospheric Parameters and Spectrum Synthesis Analysis}

\subsection{Stellar Parameters}

The adopted atmospheric parameters for Kepler-138 and Kepler-186 are given in Table 1. The effective temperatures adopted in this study were derived from the photometric calibrations for M-dwarfs by Mann et al. (2015) for the $V$-$J$ and, $r$-$J$ colors. The effective temperatures obtained for the two colors were very similar for the two stars, with  a mean effective temperature and standard deviation of: $T_{\rm eff}$(Kepler-138) = 3835 $\pm$ 21 K and $T_{\rm eff}$(Kepler-186) = 3852 $\pm$ 20 K, confirming that they are both early-type M-dwarfs. Note that no interstellar reddening corrections were applied, as the M-dwarfs studied here have distances less than $\sim$ 150 pc.

Comparisons were made with two other M-dwarf photometric calibrations from Casagrande et al. (2008) and Boyajian et al. (2012), both for the color indices $V$-$J$, $V$-$H$, $V$-$Ks$. The effective temperatures obtained using the Boyajian et al. (2012) calibration agreed well, but were slightly cooler ($\sim$ 30 K) than those from Mann et al. (2015), while the calibration from Casagrande et al. (2008) resulted in lower $T_{\rm eff}$'s by about $\sim$ 80 K. 

Surface gravities (log $g$) for the targets were calculated using the Bean et al. (2006) empirical relation of log $g$ as a function of stellar mass. Stellar masses were derived using the calibrations described in Delfosse et al. (2000) for the absolute magnitudes $M_{V}$, $M_{J}$, $M_{H}$, and, $M_{K}$, which were calculated assuming stellar distances of 66.5 pc and 151 pc for Kepler-138 (Pineda et al. 2013) and Kepler-186 (Quintana et al. 2014), respectively. The derived masses are: M/M$_{\odot}$ = 0.59 $\pm$ 0.02 for Kepler-138 and M/M$_{\odot}$ = 0.52 $\pm$ 0.03 for Kepler-186 (mean masses and standard deviations calculated using the four adopted absolute magnitudes) and the derived values of surface gravities are: log $g$ = 4.64 dex for Kepler-138 and log $g$ = 4.73 for Kepler-186 (Table 1). 

In order to estimate the uncertainties in the derived atmospheric parameters, we used the internal errors quoted in the studies mentioned above: Mann et al. (2015) report an internal uncertainty for the $T_{\rm eff}$ calibration of $\sim$ $\pm$60 K; Delfosse et al. (2000) find their results to have a precision of 10\% which returns an uncertainty of $\sim$ $\pm$0.05 $M_{\odot}$, and Bean et al. (2006) estimate that the internal uncertainty in their calibration is $\pm$0.08 dex in log $g$. We also added the uncertainties related to the errors in the photometry of $\sim$  0.03 mags. The estimated uncertainties in $T_{\rm eff}$, log $g$ and stellar mass were added in quadrature and result in final uncertainties of $\pm$ 64 K in $T_{\rm eff}$; $\pm$ 0.06 $M_{\star}$; and  $\pm$ 0.10 dex in the log $g$ (Table 1).

In addition, estimates of the effective temperatures for the target stars could also be obtained from their spectra.  Spectroscopic $T_{\rm eff}$'s were derived from the two oxygen abundance indicators present in the APOGEE spectra of these M-dwarfs: the OH lines (Table 2) and H$_{2}$O lines (at 15253.3 \AA{}, 15315.7 \AA{}, 15317.3 \AA{} and 15334.9 \AA{}). Figure 2 shows the variation of oxygen abundance with effective temperature from the OH (solid line) and H$_{2}$O lines (dashed line) for Kepler-138. (The methodology for the abundance analysis is described in Section 3.2). The oxygen abundances displayed were computed by changing the effective temperature in the model atmospheres, while keeping the log $g$ constant (Table 1). The OH lines are rather insensitive to the effective temperature, in contrast to the H$_{2}$O lines, which show a much steeper O-abundance trend with $T_{\rm eff}$, offering the possibility of finding the $T_{\rm eff}$ - A(O) pair that best matches the spectral lines of both OH and H$_{2}$O simultaneously. The point of agreement in the oxygen abundance for Kepler-138 is $T_{\rm eff}$ = 3833 K, as shown in Figure 2, while for Kepler-186 it is $T_{\rm eff}$ = 3850 K. The spectroscopic values of $T_{\rm eff}$ derived using this method are effectively indistinguishable, within the uncertainties, from the effective temperatures computed using the Mann et al. (2015) photometric calibration. Newton et al. (2015) also derived effective temperatures for the target stars from $H$-band low-resolution spectra. Their derived $T_{\rm eff}$ for Kepler-138 ($T_{\rm eff}$ = 3841 K) agrees with our determination; for Kepler-186, however, they find a much cooler effective temperature ($T_{\rm eff}$ = 3624 K), which is not in agreement with the H$_{2}$O and OH lines in the observed APOGEE spectrum of Kepler-186. A more detailed discussion of the effective temperature scale for the APOGEE spectra will be presented in Souto et al. (2016; see also Schmidt et al. 2016). 

\subsection{Abundance Analysis}

Since most of the lines in the APOGEE spectra of M-dwarfs, both atomic and molecular, are blended to varying degrees, spectrum synthesis fitting was used, rather than equivalent-width measurements, to compute chemical abundances of several elements. 
The abundance analysis presented here is based on the LTE spectral synthesis code $turbospectrum$ (Plez 2012; Alvar\'ez \& Plez 1998) together with plane-parallel MARCS model atmospheres (Gustafsson et al. 2008) computed specifically for the stellar parameters derived above.  
As part of this 1-D analysis, we investigated the sensitivity of the spectral lines to the microturbulence parameter. The synthetic spectra exhibited little sensitivity to the microturbulent velocity for most of the spectral lines, except for the OH lines, which were found to be more sensitive. The OH lines were then used to estimate the microturbulent velocities, using a similar methodology as described in Smith et al. (2013) for Fe I. We computed oxygen abundances corresponding to $\xi$ = 0.50, 0.75, 1.00, 1.25, 1.50 km.s$^{-1}$ and the selected value of $\xi$ was the one showing the lowest spread in the oxygen abundances from the different OH lines. We obtained $\xi$ = 1.0 $\pm$ 0.25 km s$^{-1}$ for both Kepler-138 and Kepler-186.

\subsubsection{Enhancements to the APOGEE Line List for the study of M-dwarfs}

The most recent version of the APOGEE line list for DR13 (tagged version 20150714) was used in the computation of synthetic spectra. The APOGEE DR13 line list is an updated version of the DR12 line list (Shetrone et al. 2015; see also Holtzman et al. 2016). In the context of the study of cool dwarfs, an important improvement in the DR13 line list, relative to DR12, is the inclusion of the strongest water lines from Barber et al. (2006): 1,891,108 water transitions were added. Inclusion of water transitions in the APOGEE line list is crucial for modeling the spectra of M-dwarfs and, in particular, of M-dwarfs of later spectral types (Allard 2000; Tsuji et al. 2015). At the effective temperatures of the studied stars ($T_{\rm eff}$ $\sim$ 3850 K) the intensities of H$_{2}$O lines, however, are generally weak, with the stronger water lines falling in the blue part of the APOGEE spectra. Figure 3 shows in the top panels synthetic spectra computed only with the water line list for two spectral regions of the APOGEE spectra (left side between 15200 -- 15300 \AA{} and right side between 16500 -- 16600 \AA{}). The presence of weak water lines throughout the spectra has the general effect of lowering the pseudo stellar continuum by at most $\sim$ 2 percent. However, these lines of H$_{2}$O, although relatively weak in the spectra of the target stars, are quite valuable for constraining their effective temperatures, as discussed in Section 3.1.

As previously mentioned, the APOGEE line list does not include any transitions of iron hydride, an important contributor to the spectra of M-dwarfs ({\"O}nehag et al. 2012; Lindgren et al. 2016). It was clear from our test syntheses that the DR13 line list could not match well several features in the observed spectra of the target stars. To investigate if the unmatched features were due to FeH, we first tested the line list from Phillips et al. (1987) that includes the vibration-rotation transition of the FeH $\Delta^{4}$ - $\Delta^{4}$ system. The computed FeH lines from this line list were weak and did not improve the matching of the observed spectra. Therefore, the missing lines in the line list are probably not from the F-X FeH-transition and the Phillips et al. (1987) was not included in the line list for this study. The unmatched lines, however, could be from the E-X FeH transition with a bandhead at 1.35 $\mu$ (described in Balfour et al. 2004).
The addition of the FeH lines from Hargreaves et al. (2010), on the other hand, improved the matching of several of the unidentified features in the synthetic spectra. Hargreaves et al. (2010) studied the electronic transition for the FeH $E^{4}$$\Pi$ - $A^{4}$$\Pi$ system in the near-IR from 1.58 $\mu$m up to 1.7 $\mu$m, see also Wallace \& Hinkle (2001).

The gf-values for the FeH lines from Hargreaves et al. (2010) were computed using their line intensities and the expression for converting HITRAN-like intensities to Einstein A-values from {\v S}ime{\v c}kov{\'a et al. (2006), equation (20). The A-values were converted to gf-values using the standard expression from Larsson (1983):

g$f$ = [1.499(2J+1)A].[($\sigma$$^{2}$)$^{-1}$], $A$-value is the Einstein A coefficient, $J$ is the lower state angular momentum and $\sigma$ is the wave number.

Synthetic spectra computed only with the FeH lines are shown in the second top-to-bottom panels of Figure 3. Note the absence of FeH transitions in the region between 15200 -- 15300 \AA{} (left panel), due to the fact that the FeH lines do not extend blueward of the band-head at 15820 \AA{}. In the right panel (16500 -- 16600 \AA{}), conspicuous FeH lines are seen, with several lines reaching depths of 10\% relative to the continuum. 

Additional line lists for the hydrides MgH, CaH, and CrH from Kurucz (1993) were also tested, but these did not produce improvements in the overall fits, nor could we identify any matching features from these hydrides in the target star spectra. We note that there remain a few unidentified features, e.g., at 15232 \AA{} and 16059 \AA{}. Future work on the line list is still needed in order to fully match the spectra of M-dwarfs in the APOGEE region. The line list constructed for this study is however, fairly complete in order to derive detailed chemical abundances of several elements. The third row (top-to-bottom) panels show the DR13 line list including atomic and molecular lines. Best fitting spectra of Kepler-186 using the enhanced line list for this study are illustrated in the bottom panels of the Figure 3.

\subsection{Derived Abundances}
We derived the chemical abundances of thirteen elements: C, O, Na, Mg, Al, Si, K, Ca, Ti, V, Cr, Mn, and Fe in the APOGEE spectra of the M-dwarfs Kepler-138 and Kepler-186. Table 2 lists the selected transitions for this manual abundance analysis, as well as, the excitation potentials (eV) and oscillator strengths (log $g$f) of the lines from DR13, except for FeH (see 3.2.1). The selected lines for the abundance analyses are also indicated in Figure 1 (with dots underneath the features). These spectra are dominated by a large number of OH lines (more than 70 OH lines are easily identified in the spectra). 

Unfortunately, lines of the CN molecule, as well as the atomic lines of S I, Co I, Ni I, and Cu I, which can be measured in the APOGEE spectra of red-giants (Smith et al. 2013), become very weak and mostly blended with other species in the M-dwarfs.

The abundances for most of the studied elements were derived from neutral atomic lines, with the exception of C and O, for which we used molecular lines of $^{16}$OH (36 individual transitions of OH) and $^{12}$C$^{16}$O (four regions containing CO lines $\lambda$15570--15600, $\lambda$15970--16010, $\lambda$16182--16186, $\lambda$16600--16650), respectively. We also checked the possibility of measuring carbon abundances from the C I lines (at 15784.546 \AA{}, 16004.900 \AA{} and 16890.417 \AA{}) but found these lines to be too weak. We derived a carbon abundance from the weak CO lines (for an assumed oxygen abundance, although the CO lines are not very sensitive to the oxygen abundances) and then derived an oxygen abundance from the various strong OH lines. In principle, there is one combination of carbon and oxygen abundances that will produce a good match for both the CO and OH lines, as the OH lines are also sensitive to the carbon abundances. As discussed above, the spectra do not show any contribution from CN and the synthetic spectra were very insensitive to any changes in the nitrogen abundances.

Best fitting synthetic spectra were obtained both from visual comparisons of the observed and modeled spectra around the selected lines for abundance analyses and from the computed differences and standard deviations between the observed spectrum and synthesis. We manually defined the windows covering the selected lines, applied any needed adjustments in radial velocities and defined the pseudo-continuum either locally or in more distant regions depending on each case.  Once we had a visual best fit abundance we computed the minimum  $\chi$$^{2}$ and verified that the best abundance corresponded to a minimum $\chi$$^{2}$ in each case. The synthetic spectra were broadened with a Gaussian profile corresponding to a full-width half maximum (FWHM $\sim$ 0.73 \AA{}), given by the APOGEE spectral resolution. We note that in some cases Gaussian profiles with slightly different FWHM (from $\sim$ 0.65 -- 0.8 \AA) were used to adequately fit the observed line profiles. This is because the APOGEE LSF and the resolution vary slightly depending on the fiber and spectral region (Holtzman et al. 2015). The target M-dwarfs are rotating slowly and below the limit set by the resolution of the instrumental profile. No extra broadening beyond the FWHM was needed to fit the spectra, although marginally larger FWHM were needed to fit Kepler-186 when compared to Kepler-138. Figure 3 (bottom panels) shows best fitting spectra obtained for the target star Kepler-138. Detailed examples of the fits obtained for at least one spectral line per studied element for both Kepler-138 (blue spectra) and Kepler-186 (green spectra) are shown in Figure 4.
Individual line and molecular feature abundances are presented in Table 2, along with mean values and their standard deviations. 

The iron abundances were derived from the sample Fe I lines in Table 2. There are, however, systematic differences between the mean iron abundances based on Fe I transitions and the iron abundances from best overall fits obtained for the FeH lines in the observed M-dwarf spectra: the FeH lines indicate an iron abundance on average 0.10 - 0.15 dex lower than the mean value using the sample Fe I lines (both values for the Fe abundances are listed in Table 2). The gf-values for the Fe I transitions in this study are from DR13 and have been adjusted to fit the spectra of the Sun and Arcturus; adjustments were only allowed within 2$\sigma$ of the gf-value estimated uncertainties (Shetrone et al. 2015). The gf-values for the FeH transitions were computed from their intensities in Hargreaves et al. (2010; Section 3.2.1); such gf-values may be more uncertain and have not yet been verified. It is impossible to derive astrophysical gf-values for FeH lines using the same scheme as for the construction of the DR13 line list, as FeH is not measurable in the spectra of the Sun and Arcturus. Due to the uncertainties in the FeH gf-values, we adopt iron abundances based upon the Fe I lines in this study. Improvements to the FeH gf-values will be presented in a future paper (Souto et al. 2016). 

The uncertainties of the derived abundances due to uncertainties in the adopted stellar parameters can be estimated in a manner similar to that presented in Souto et al. (2016). We adopted the model atmosphere used in the analysis of Kepler-138 as a baseline model, and changed the atmospheric parameters by $T_{\rm eff}$ + 65 K; log $g$ + 0.10 dex; [M/H] + 0.20 dex; C/O + 0.15, one at a time (Table 3). The sensitivity to the microturbulent velocity parameter is also given ($\xi$ + 0.25 km s$^{-1}$) but it is negligible for all species, except OH. The last column in Table 3 presents the quadrature sum of abundance changes due to variations in the parameters as discussed above. Note that while the errors in [Mg/H] and [Si/H] are dominated by uncertainties in the effective temperature, these errors effectively cancel out in the ratio [Mg/Si]. We also estimate an abundance uncertainty of $\sim$ 0.02 dex and $\sim$ 0.06 dex that comes from overall uncertainties in setting the pseudo-continuum in the spectra of Kepler-138 and Kepler-186, respectively. These uncertainties in the pseudo-continuum placement are folded into the final abundance uncertainties which are presented in Table 4, together with the mean abundances (in the [X/H] notation) for the two studied stars.

\subsection{Comparisons with Metallicities Derived from Low-Resolution Spectra}

The abundances from Table 2 show that the two exoplanet-hosting M-dwarfs Kepler-138 and Kepler-186 are slightly metal-poor and have very similar mean Fe abundances and standard deviations: $\langle$A(Fe)$\rangle$$_{\rm Kepler-138}$ = 7.36 $\pm$ 0.05 ($\pm$ 0.09)  and $\langle$A(Fe)$\rangle$$_{\rm Kepler-186}$ = 7.37 $\pm$ 0.06 ($\pm$ 0.10); the numbers in parentheses represent the total estimated abundance uncertainties. These results (based on measurements of Fe I lines in the APOGEE spectra) corresponding to [Fe/H]$_{\rm Kepler-138}$ = -0.09 dex and [Fe/H]$_{\rm Kepler-186}$ = -0.08 dex, are higher than the iron abundances obtained previously by Muirhead et al. (2014) for these stars, based on low-resolution $K$-band spectra from the TripleSpec spectrograph on the Palomar Hale-5m telescope. The latter study obtained: [Fe/H]$_{\rm Kepler-138}$ = -0.25 $\pm$ 0.12 and [Fe/H]$_{\rm Kepler-186}$ = -0.34 $\pm$ 0.12; therefore, finding both stars to be somewhat more metal-poor than the result obtained with the APOGEE spectra. To estimate [Fe/H], Muirhead et al. (2014) use the same technique as described in Rojas-Ayala et al. (2012), based on equivalent width measurements of the Na I doublet and Ca I triplet lines as well as the H$_{2}$O-K2 index (Covey et al. 2010) as a $T_{\rm eff}$ indicator. In addition, Muirhead et al. (2014; Table 1 in their paper) computed metallicities using two other calibrations involving infrared spectroscopic indices in the $H$-band by Terrien et al. (2012; [Fe/H]$_{\rm Kepler-138}$ = -0.25 $\pm$ 0.13 and [Fe/H]$_{\rm Kepler-186}$ = -0.20 $\pm$ 0.11) and in the $K$-band by Mann et al. (2013; [Fe/H]$_{\rm Kepler-138}$ = -0.22 $\pm$ 0.12 and [Fe/H]$_{\rm Kepler-186}$ = -0.30 $\pm$ 0.13. 
The general conclusion from their comparisons is that the methods overall agree but with some scatter. 
Terrien et al. (2015b) also obtained [Fe/H] = -0.21 for Kepler-138 based on $H$-band spectra. These sets of metallicity results, all from low-resolution spectra and not based on transitions involving Fe itself, are roughly consistent with each other but, on average, more metal-poor by about $\sim$0.2 dex than what is obtained here from the detailed analysis of Fe I lines here. 

In a recent study that explores more deeply the low-resolution metallicity analysis technique, Veyette et al. (2016) investigated the impact of varying the C and O abundances on the metallicities derived from calibrations that are based on spectral indices measured from low-resolution spectra (Rojas-Ayala et al. 2012; Terrien et al. 2012; Mann et al 2013; Newton et al. 2014). Veyette et al. (2016) use synthetic spectra generated with PHOENIX models (Allard et al. 2012a,b) to evaluate how the C/O ratio influences the measured equivalent widths of the Na I and Ca I features that are used to determine metallicities. They find that the pseudo-continua of synthetic spectra of M-dwarfs are very sensitive to the C/O abundance ratios. In essentially all M-dwarfs the carbon-to-oxygen abundance ratio is less than 1 (C/O$<$1) and the very stable CO molecule binds nearly all C into CO, with the remaining O atoms (defined by O-C) going into H$_{2}$O and OH.  As Veyette et al. (2016) point out, CO and H$_{2}$O dominate the line opacity of M-dwarfs in the near-IR, and the strengths of the CO and H$_{2}$O lines are sensitive functions of the C/O ratio.  The molecular absorption from these species thus define the pseudo-continuum against which the Na I and Ca I equivalent widths are defined.  
The detailed modelling in Veyette et al. (2016) demonstrates the importance of the C/O ratios (or, as they point out, the crucial variable is O-C), because differences greater than 1 dex in [Fe/H] can be obtained based on different assumed values of C/O when using the Na I and Ca I indicators. 
The derivation of individual, precise carbon and oxygen abundances, such as is presented in this analysis of high-resolution APOGEE spectra, may help in calibrating some of the low-resolution metallicity indicators for M-dwarfs. This will be investigated in analyses of much larger samples of M-dwarfs (Souto et al. 2016) and will be particularly interesting for the coolest dwarfs.

\section{Detailed Chemical Abundance Distributions}

The near-IR APOGEE spectra are shown here enable the derivation of detailed abundance patterns in M-dwarf stars; this was discussed and demonstrated in Sections 3.2 and 3.3, with abundances of thirteen elements presented in Tables 2 and 3.
The mild Fe-deficiency of $\sim$ -0.1 dex, relative to the Sun is consistent with the other elemental abundances, with global means of all elements measured being [M/H] = -0.13 $\pm$ 0.06 ($\pm$ 0.10) for Kepler-138 and [M/H] = -0.08 $\pm$ 0.10 ($\pm$ 0.10) for Kepler-186, where M is the mean of all elemental abundances. Looking more closely at nucleosynthesis patterns, the average of $\alpha$-element abundances (where [$\alpha$/Fe] = [(Mg+Si+Ca+Ti/4)/Fe]) in Kepler-138 is ([$\alpha$/Fe] = -0.02 dex) and Kepler-186 ([$\alpha$/Fe] = +0.06 dex), which suggests a slight $\alpha$-element enhancement in Kepler-186, especially for silicon and, to a lesser extent, magnesium. 
The average odd-Z elemental abundances ([(Na+Al+K/3)/Fe])$_{\rm Kepler-138}$ = [(Na+Al+K/3)/Fe]$_{\rm Kepler-186}$ = -0.07) and the iron-peak elements ([(Cr+Mn/2)/Fe]$_{\rm Kepler-138}$ = 0.01 dex, [(Cr+Mn/2)/Fe]$_{\rm Kepler-186}$ = 0.02 dex) show a behavior very similar to the Galactic trends for thin disk stars in the Solar neighborhood (e.g., Adibekyan et al. 2012; Bensby et al. 2014).
The abundance ratio of carbon-to-oxygen also follows Galactic trends, with derived values of C/O = 0.55 $\pm$ 0.10 and C/O = 0.52 $\pm$ 0.12, respectively, for Kepler-138 and Kepler-186, or, a nearly solar C/O ratio (Asplund et al. 2005; see also Caffau et al. 2011).  The C/O ratios for both Kepler-138 and Kepler-186 fall on the trend of C/O versus [Fe/H] defined in Nissen et al. (2014).

The slight enhancement of silicon in Kepler-186, with [Si/Fe] = +0.18, could be of interest as Si is an important element in determining the internal structure of Earth-sized rocky planets (e.g., Unterborn, Dismukes, \& Panero 2016; Brewster \& Fischer 2016). It is worth noting that the Si I lines sample both weak and strong lines, so the Si abundance is likely to be fairly secure. By itself, the small enhancement of Si relative to Fe in Kepler-186 may not be significant, in particular because the APOGEE spectrum of Kepler-186 is noisier than the one for Kepler-138. Larger samples of exoplanet-hosting M-dwarfs need to be studied in high-resolution to further investigate if this is a real signature. (See also Terrien et al. 2015b for potential methods for estimating $\alpha$/Fe from low-resolution spectra).

The recent studies of Unterborn \& Panero (2016) and Brewer \& Fischer (2016) point out the importance of specific key elements in affecting the interior structures of small planets, with the examples in these papers being the elemental ratios of C/O and Mg/Si. (See also Dorn et al. 2015 and Alibert 2016). Figure 5 shows results for linear values of C/O versus Mg/Si for the sample of 849 stars from Brewer \& Fischer (2016; see also Delgado-Mena et al. 2010), which contains both known exoplanet-hosting FGK-dwarfs, as well as FGK-dwarfs whose exoplanet-hosting status is unknown. Given the results from the Kepler mission, as well as modeling of those results, it is likely that nearly all FGK-dwarfs host small (super Earth-sized) planets; thus a well-defined criterion of `exoplanet host' or `non-exoplanet host' may not be possible to apply.  Perhaps a better label would be to associate known exoplanet-hosting stars with the type of exoplanets that they host.  Also shown are in Figure 5 are results from Schuler et al. (2015) for five Kepler G-dwarfs that host small exoplanets that are similar to the types of exoplanets hosted by Kepler-138 and Kepler-186; estimated uncertainties are shown as error bars for the two M-dwarfs analyzed here. As mentioned, the plotted quantities are on a linear scale and the scatter from the results in Brewer \& Fischer (2016) and Schuler et al. (2015) is quite small. Both Kepler-138 and Kepler-186 fall within the scatter defined by the much hotter FGK-dwarfs, with the already discussed somewhat high Si abundance in Kepler-186 pulling its Mg/Si value to the low side of the distributions.  Chemical evolution in this diagram moves from larger Mg/Si and smaller C/O ratios at lower metallicity and moves to lower Mg/Si and larger C/O ratios as metallicity increases (e.g., Nissen et al. 2014), or from lower right to upper left.  Being near-solar metallicity, both Kepler-138 and Kepler-186 exhibit roughly solar values of C/O, with Kepler-138 also having nearly the same Mg/Si ratio as the Sun, but with Kepler-186 slightly off the main trend because of its slight Si enhancement.  

\section {Summary}

The main result from this study is to demonstrate that a detailed chemical abundance analysis of the elements C, O, Na, Mg, Al, Si, K, Ca, Ti, V, Cr, Mn and Fe can be derived high-resolution 1.5 -- 1.7 $\micron$ APOGEE spectra of early M-dwarfs. This is the first study of this type and the first time that detailed chemical abundances of a large number of elements are presented for M-dwarfs in general, and for Kepler exoplanet-hosting M-dwarfs, in particular. 

The target stars are interesting early type M-dwarfs that host Earth-sized, or smaller, exoplanets, one of which falls in the habitable zone and the other having about the size of Mars.  Both of these low-mass stars are found to be slightly metal-poor, with near-solar C/O ratios, and other abundance ratios that are also near-solar. Kepler-186 exhibits a marginally enhanced [Si/Fe] ratio; silicon is an important element when modelling the interiors of rocky exoplanets. The average of the $\alpha$-element abundances, relative to Fe, in Kepler-186 (O, Mg, Si, Ca, Ti) is [$\alpha$/Fe] = +0.05, which is what would be expected for a slightly metal-poor star.

Careful comparisons with metallicities derived from low-resolution spectral measurements in M-dwarfs indicate possible systematic differences of $\sim$ 0.2 dex  and future comparisons using larger samples of APOGEE spectra may aid in calibrations for low-resolution spectra of M-dwarfs.

\acknowledgments
\section{Acknowledgments}
We warmly thank Bertrand Plez for pointing out the Hargreaves et al. (2010) study that was fundamental for constructing the line list for this study. We thank the anonymous referee for useful comments that helped improve the paper.

Funding for the Sloan Digital Sky Survey IV has been provided by
the Alfred P. Sloan Foundation, the U.S. Department of Energy Office of
Science, and the Participating Institutions. SDSS-IV acknowledges
support and resources from the Center for High-Performance Computing at
the University of Utah. The SDSS web site is www.sdss.org.

SDSS-IV is managed by the Astrophysical Research Consortium for the 
Participating Institutions of the SDSS Collaboration including the 
Brazilian Participation Group, the Carnegie Institution for Science, 
Carnegie Mellon University, the Chilean Participation Group, the French Participation Group, Harvard-Smithsonian Center for Astrophysics, 
Instituto de Astrof\'isica de Canarias, The Johns Hopkins University, 
Kavli Institute for the Physics and Mathematics of the Universe (IPMU) / 
University of Tokyo, Lawrence Berkeley National Laboratory, 
Leibniz Institut f\"ur Astrophysik Potsdam (AIP),  
Max-Planck-Institut f\"ur Astronomie (MPIA Heidelberg), 
Max-Planck-Institut f\"ur Astrophysik (MPA Garching), 
Max-Planck-Institut f\"ur Extraterrestrische Physik (MPE), 
National Astronomical Observatory of China, New Mexico State University, 
New York University, University of Notre Dame, 
Observat\'orio Nacional / MCTI, The Ohio State University, 
Pennsylvania State University, Shanghai Astronomical Observatory, 
United Kingdom Participation Group,
Universidad Nacional Aut\'onoma de M\'exico, University of Arizona, 
University of Colorado Boulder, University of Oxford, University of Portsmouth, 
University of Utah, University of Virginia, University of Washington, University of Wisconsin, 
Vanderbilt University, and Yale University. D.A.G.H. was funded by the Ram\'oon y Cajal fellowship number  RYC-2013-14182. D.A.G.H. and O.Z. acknowledge support  provided  by  the  Spanish  Ministry  of  Economy  and  Competitiveness (MINECO) under grant AYA-2014-58082-P. H.J. acknowledge support from the Sven and Dagmar Sal\'{e}n foundation. SRM acknowledges support from NSF grant AST-1616684.
%{\it Facilities:} \facility{}

\clearpage

\newpage
\begin{figure}
\epsscale{1}
\includegraphics[width=7.6in,angle=-90]{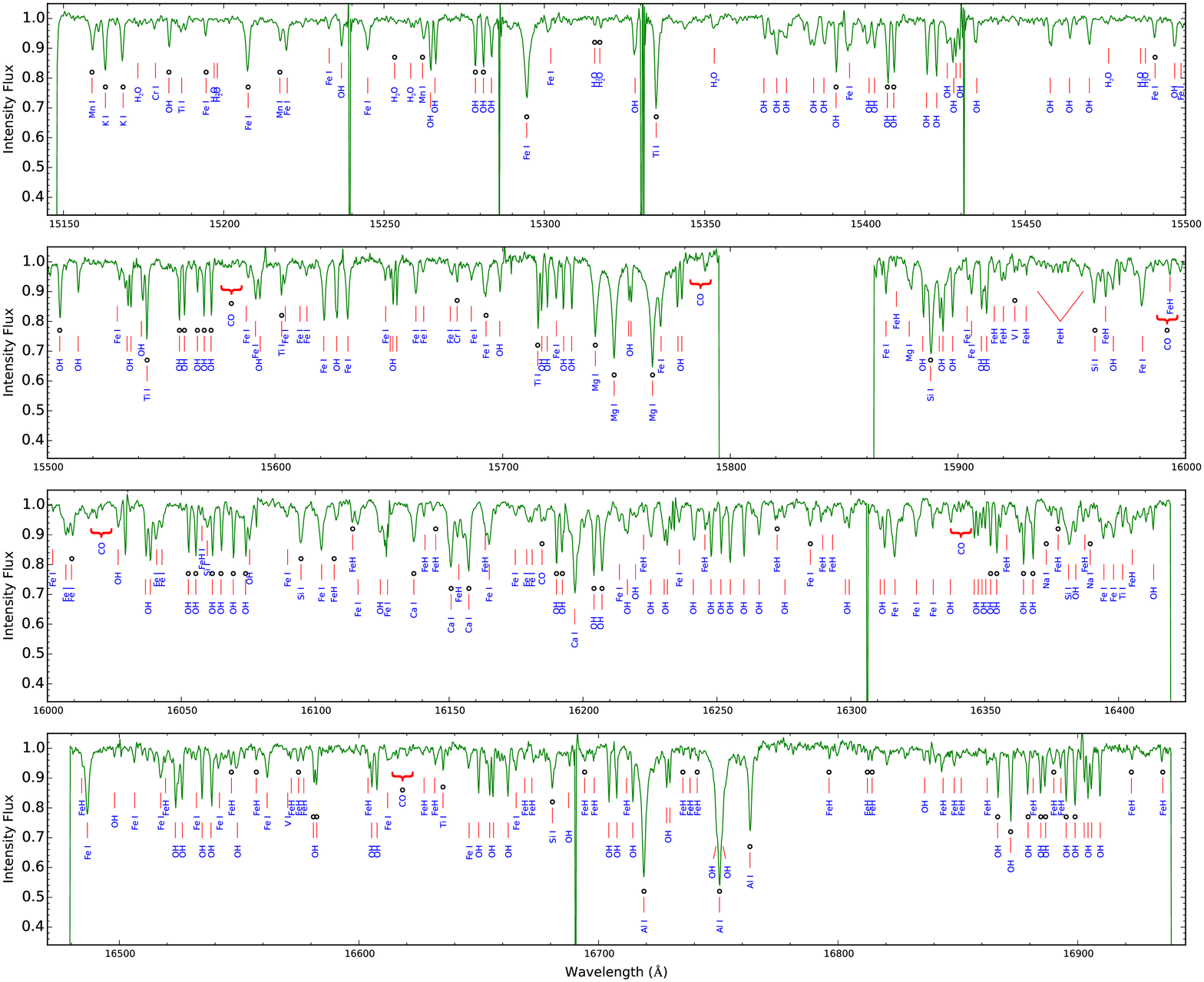}
\caption{Line identification in the APOGEE spectra of Kepler-138 corresponding to the entire APOGEE region. The lines used in the abundance analysis are indicated by the black dots. We note that most of these features are blended to some extent.}	
\end{figure}

\newpage
\begin{figure}
\epsscale{0.6}
\plotone{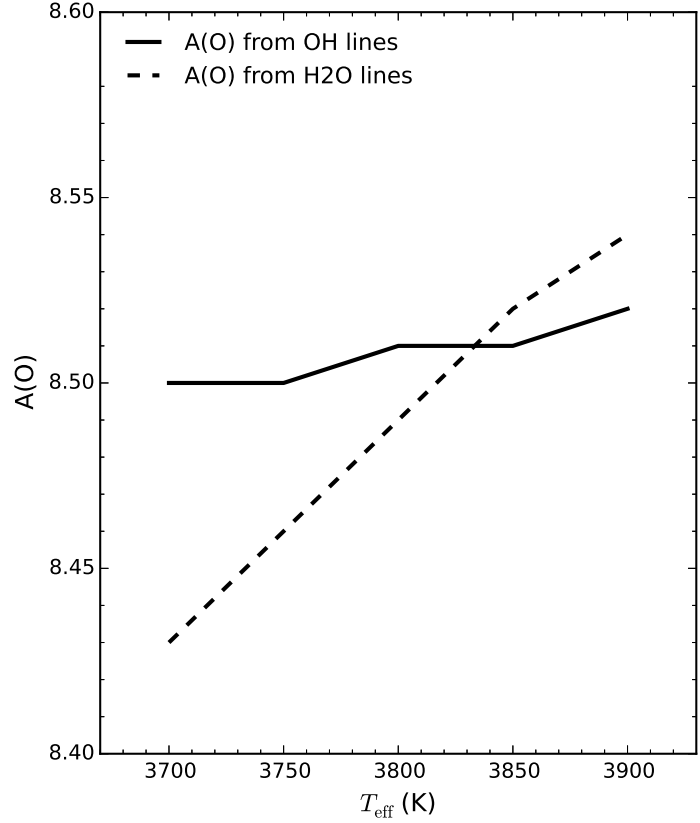}
\caption{Spectroscopic estimation of the effective temperature for Kepler-138 via the use of OH and H$_{2}$O lines. The dashed line represents the derived oxygen abundances from H$_{2}$O lines for varying the model atmosphere effective temperatures. The solid line is the same but for derived oxygen abundances from OH lines. The best $T_{\rm eff}$ is the one that brings the oxygen abundances from the two indicators into agreement. All oxygen abundance calculations were done for the same log $g$ = 4.64 as discussed in Section 3.1.}
\end{figure}

\newpage
\begin{figure}
\epsscale{0.95}
\plotone{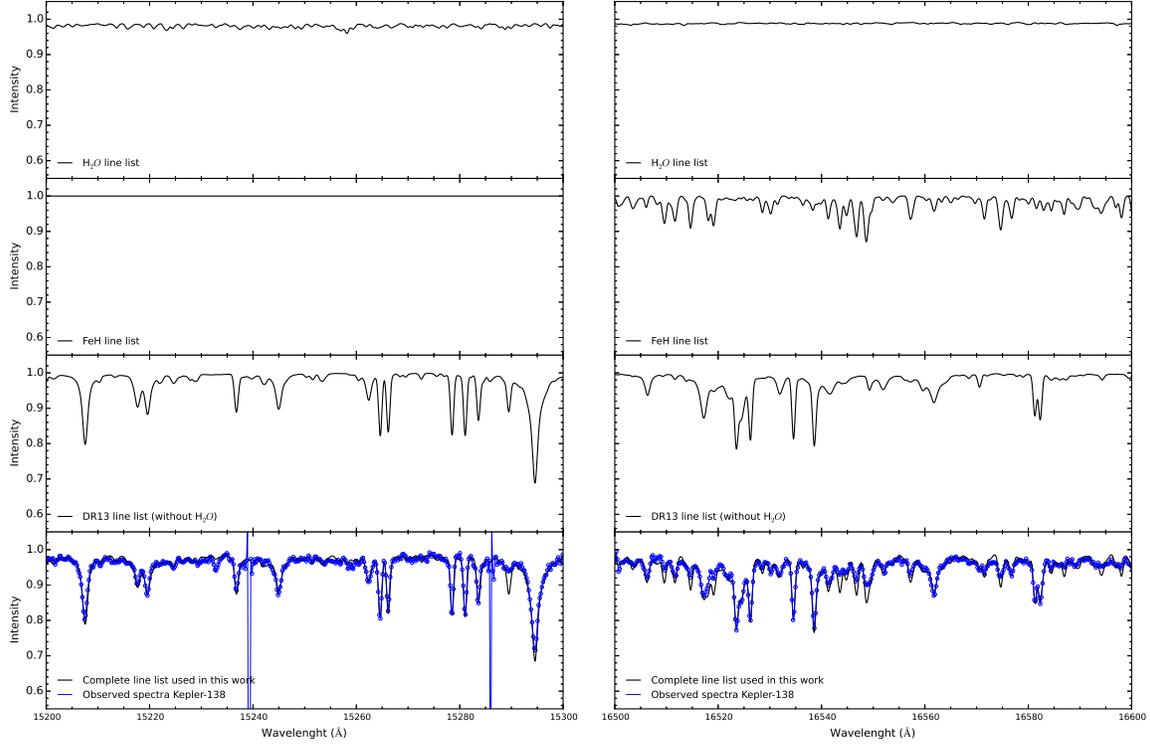}
\caption{Synthetic spectra computed using from different subsets lines of the APOGEE line list for Kepler-138 parameters: $T_{\rm eff}$ = 3835K; log $g$ = 4.64. Top panels: synthetic spectra computed using only the water line list. Second row from top to bottom: synthetic spectra computed using only the FeH line list. Third row from top to bottom: synthetic spectra computed with the DR13 line list, but with the water lines removed from the line list (DR13 line list does not include FeH). Bottom panels: Synthetic spectra computed with the modified version of the DR13 line list, which is adopted in the study, plotted with the observed spectrum of Kepler-138.}	
\end{figure}

\newpage
\begin{figure}
\epsscale{0.7}
\plotone{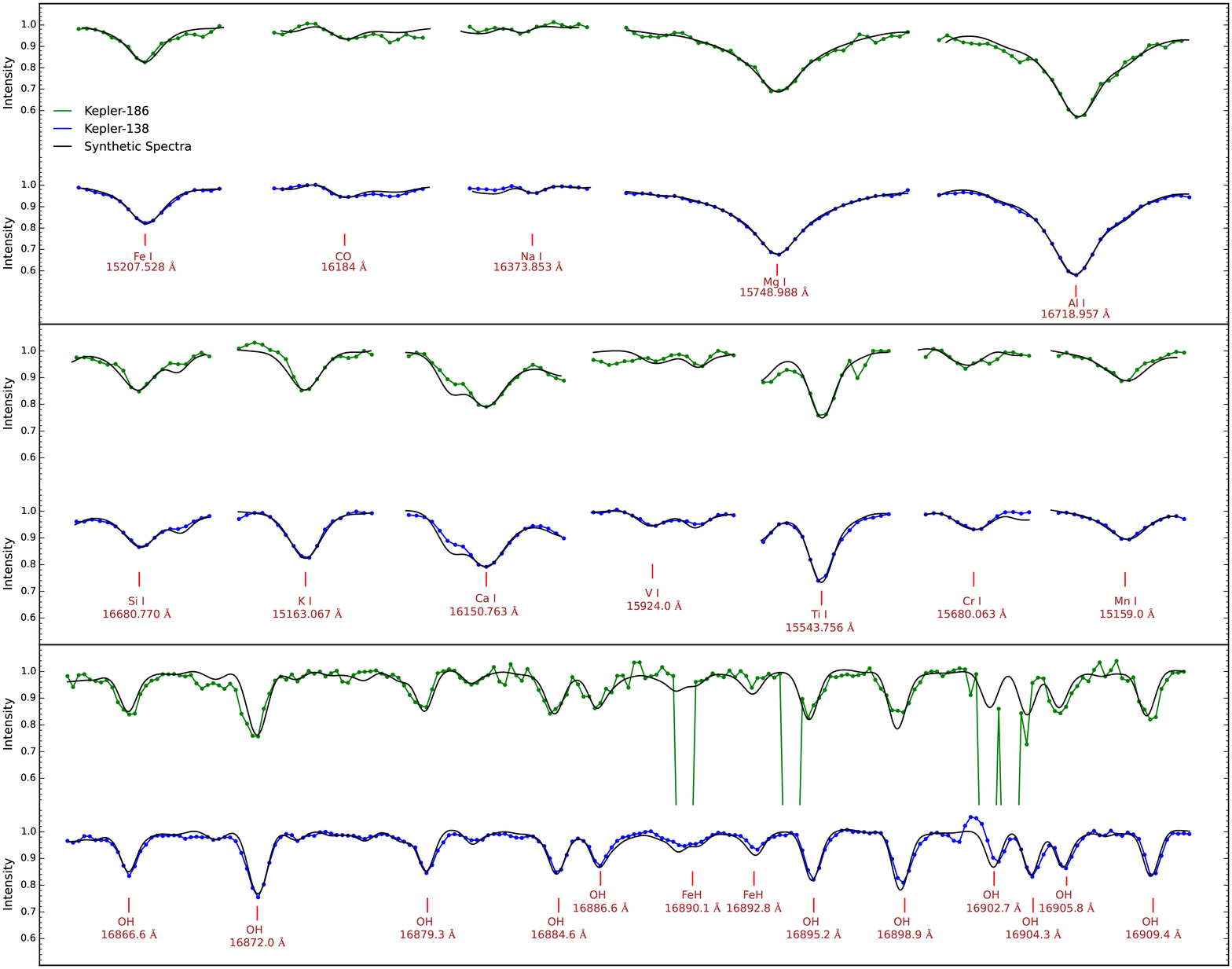}
\caption{Portions of observed spectra of Kepler-138 (blue) and Kepler-186 (green) in selected spectral regions showing at least one individual line analyzed per studied element (wavelengths are indicated). Best fitting synthetic spectra are shown as black lines. The bottom panel shows several OH lines in the spectral region between 16860 -- 16910 \AA{}.}	
\end{figure}

\newpage
\begin{figure}
\epsscale{1}
\plotone{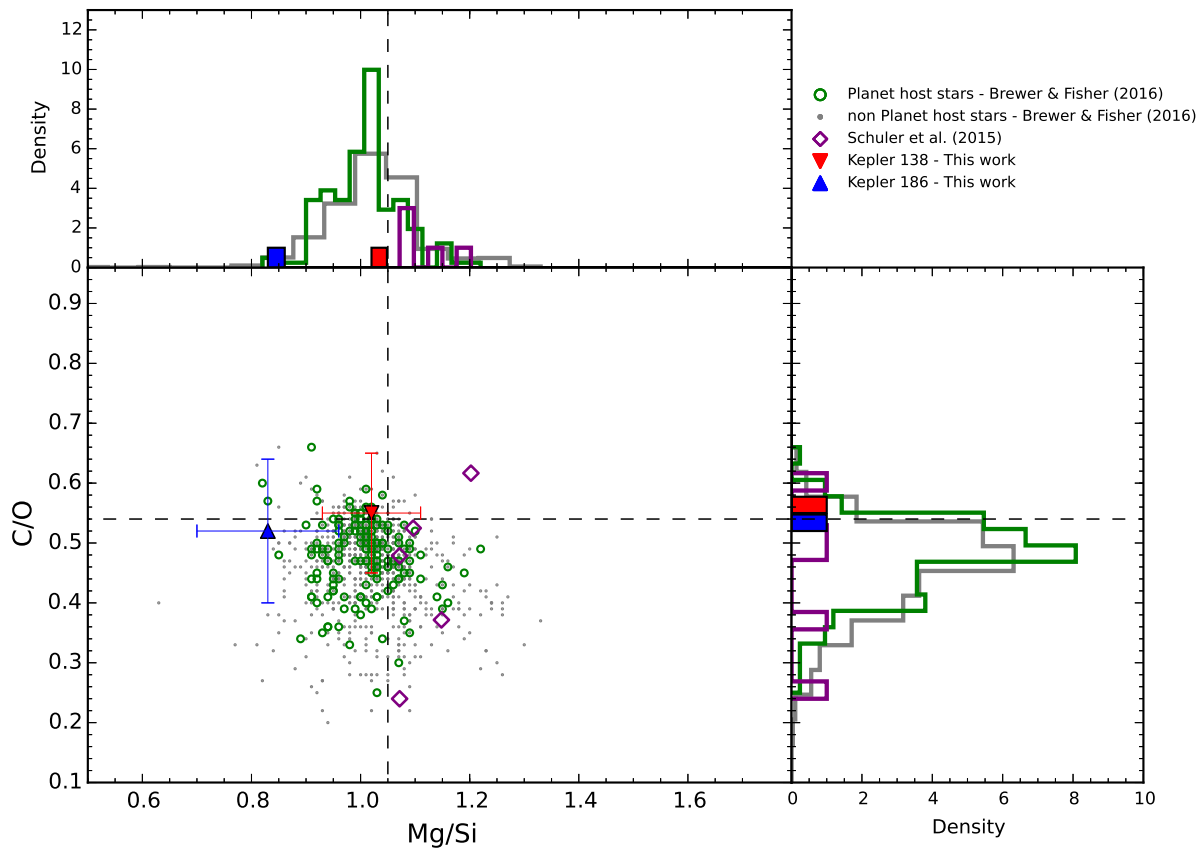}
\caption{The C/O versus Mg/Si diagram. Our results for Kepler-138 and Kepler-186 are shown as blue and red triangles. The dashed lines represent the solar values for the C/O and Mg/Si ratios. The green circles are planet-hosting stars and grey dots refer to stars with unknown planet hosting status from Brewer \& Fisher (2016), while purple diamonds are Kepler stars with small planets from Schuler et al. (2015). The top panel shows a histogram of the normalized distribution of Mg/Si; likewise in the right panel for C/O.}	
\end{figure}

\clearpage

\begin{deluxetable}{lll}
\tabletypesize{\scriptsize}
%\rotate
\tablecaption{Adopted Photometry and Atmospheric Parameters}
\tablewidth{0pt}
\tablehead{
\colhead{} &
\colhead{Kepler-138} &
\colhead{Kepler-186} 
}
\startdata
$V$				&	13.168			&	15.290\\
$J$				&	10.293			&	12.473\\
$H$				&	9.680			&	11.824\\
$Ks$			&	9.506			&	11.605\\
$r$				&	12.529			&	14.664\\
$d$ $(pc)$	&	66.5			&	151.0\\
$T_{\rm eff}$ (K)	&	3835 $\pm$ 64	&	3852 $\pm$ 64\\
log $g$		&	4.64 $\pm$ 0.10	&	4.73 $\pm$ 0.10\\
M/M$_{\odot}$		&	0.59 $\pm$ 0.06	&	0.52 $\pm$ 0.06\\
$\xi$ ($km/s$)	&	1.00 $\pm$ 0.25		&	1.00 $\pm$ 0.25

\enddata
\tablecomments{The $V$ and $r$ magnitudes were taken from UCAC4 (Zacharias et al. 2012) and the $J$, $H$, and $K_{s}$ magnitudes were taken from 2MASS (Cutri et al. 2003).}
\end{deluxetable}

\begin{deluxetable}{cccccc}
\tabletypesize{\tiny}
%\rotate
\tablecaption{Stellar Abundances}
\tablewidth{0pt}
\tablehead{
\colhead{Element} &
\colhead{$\lambda$ (\AA{})} &
\colhead{eV} &
\colhead{log $g$$f$} &
\colhead{A(Kepler-138)} &
\colhead{A(Kepler-186)}}

\startdata

{\bf Fe I}   	&15194.492	&2.223	&-4.748	&7.26	&7.27 \\
				&15207.530	&5.385	&0.138 	&7.39	&7.42 \\
                &15294.562	&5.308  &0.680	&7.30	&7.31 \\                
				&15490.339	&2.198	&-4.755	&7.30	&7.36 \\
				&15648.515	&5.426	&-0.689	&7.37	&7.45 \\
				&15692.751	&5.385	&-0.610	&7.39	&7.38 \\
				&16009.615	&5.426	&-0.556	&7.43	&7.40 \\

$\langle$A(Fe)$\rangle$ $\pm$ $\sigma$ & & & & 7.36 $\pm$ 0.05&	7.37 $\pm$ 0.06 \\
& & & & & \\
{\bf FeH}   	&16107.085	&0.279	&-1.688	&7.28	&7.25\\
				&16114.049	&0.279	&-1.282	&7.28	&7.22\\
				&16245.746	&0.142	&-1.409	&7.23	&...\\
				&16271.777	&0.302	&-1.136	&7.18	&7.26\\
				&16284.665	&0.229	&-1.274	&7.31	&7.29\\
%				&16288.906	&0.466	&-0.837	&bad	&bad\\
				&16377.403	&0.344	&-1.203	&7.22	&7.29\\
				&16546.755	&0.189	&-1.512	&7.16	&...\\
				&16557.238	&0.279	&-1.586	&7.31	&...\\
				&16574.751	&0.473	&-0.996	&7.16	&7.31\\
%				&16631.793	&0.279	&-1.521	&bad	&bad\\
				&16694.389	&0.279	&-1.485	&7.18	&...\\
				&16735.420	&0.220	&-1.326	&7.15	&...\\
%				&16738.294	&0.279	&-1.048	&bad	&bad\\
				&16741.657	&0.165	&-1.341	&7.14	&...\\
				&16796.382	&0.279	&-1.031	&7.10	&...\\
				&16812.687	&0.279	&-1.041	&7.13	&7.24\\
				&16814.063	&0.178	&-1.277	&7.17	&7.20\\
				&16889.575	&0.194	&-1.239	&7.20	&...\\
				&16892.878	&0.279	&-1.055	&7.21	&...\\
				&16922.746	&0.251	&-1.259	&7.23	&...\\
				&16935.090	&0.293	&-1.254	&7.11	&7.31\\

$\langle$A(Fe)$\rangle$ $\pm$ $\sigma$ & & & & 7.20 $\pm$ 0.06&	7.26 $\pm$ 0.04 \\
& & & & & \\

{\bf CO}  		&15570-15600&		&		&8.26	&8.22\\
				&15970-16010&		&		&8.25	&8.35\\
				&16182-16186&		&		&8.21	&8.22\\
				&16600-16650&		&		&8.22	&8.32\\
$\langle$A(C)$\rangle$ $\pm$ $\sigma$ & & & & 8.24 $\pm$ 0.02&	8.30 $\pm$ 0.03 \\
& & & & & \\
{\bf OH}  		&15183.943	&4.757	&-10.703&8.51	&8.55	\\
				&15278.334	&3.852	&-8.499	&8.53	&8.56	\\
				&15280.884	&1.69	&-5.855	&8.51	&8.50	\\
				&15283.771	&0.494	&-7.786	&8.52	&8.56	\\
				&15391.208	&0.494	&-5.437	&8.53	&8.57	\\
				&15407.288	&0.255	&-5.365	&8.53	&8.58	\\
				&15409.308	&4.329	&-8.532	&8.53	&8.60	\\
				&15505.782	&1.888	&-6.177	&8.50	&8.55	\\
				&15558.023	&0.304	&-5.308	&8.53	&8.61	\\
				&15560.244	&0.304	&-5.300	&8.51	&8.61	\\
				&15565.961	&2.783	&-4.699	&8.50	&8.58	\\
				&15568.780	&0.299	&-5.269	&8.53	&8.61	\\
				&15572.084	&0.300	&-5.269	&8.51	&...	\\
				&16052.765	&0.639	&-4.910	&8.50	&8.58	\\
				&16055.464	&0.640	&-4.910	&8.51	&...	\\
				&16061.700	&0.476	&-5.159	&8.49	&8.58	\\
				&16065.054	&0.477	&-5.159	&8.49	&8.57	\\
				&16069.524	&0.472	&-5.128	&8.49	&8.56	\\
				&16074.163	&0.473	&-5.128	&8.42	&...	\\
				&16190.263	&0.915	&-5.145	&8.55	&8.61	\\
				&16192.208	&3.508	&-7.471	&8.51	&8.60	\\
				&16204.076	&0.683	&-4.851 &8.46	&8.64	\\
				&16207.186	&0.683	&-4.851	&8.48	&8.66	\\
				&16352.217	&0.735	&-4.835	&8.52	&...	\\
				&16354.582	&0.735	&-4.835	&...	&...	\\
				&16364.590	&0.730	&-4.796	&8.57	&8.59	\\
				&16368.135	&0.731	&-4.797	&8.51	&8.59	\\
				&16581.250	&3.200	&-5.754	&8.51	&8.59	\\
				&16582.013	&3.197	&-5.744	&8.53	&8.59	\\
				&16866.688	&0.762	&-4.999	&8.54	&8.60	\\
				&16871.895 	&0.763	&-4.999 &8.52	&8.57	\\
				&16879.090	&0.761	&-4.975	&8.52	&8.53	\\
				&16884.530	&1.612	&-7.532	&8.48	&8.56	\\
				&16886.279	&1.059	&-4.662	&8.48	&8.56	\\
				&16895.180	&0.901	&-4.685	&8.49	&8.57	\\
				&16898.887	&1.2751	&-5.283	&8.46	&...	\\

$\langle$A(O)$\rangle$ $\pm$ $\sigma$ & & & & 8.50 $\pm$ 0.03&	8.58 $\pm$ 0.03 \\
& & & & & \\
{\bf Na I}		&16373.853	&3.753	&-1.348	&6.11	&6.16	\\
				&16388.858	&3.753	&-1.044	&6.09	&...	\\

$\langle$A(Na)$\rangle$ $\pm$ $\sigma$ & & & & 6.10 $\pm$ 0.01&	6.16 $\pm$ 0.01 \\
& & & & & \\
{\bf Mg I}		&15740.716	&5.931	&-0.312	&7.53	&7.63	\\
				&15748.988	&5.932	& 0.125	&7.42	&7.49	\\
				&15765.842	&5.933	&0.423	&7.33	&7.46	\\

$\langle$A(Mg)$\rangle$ $\pm$ $\sigma$ & & & & 7.43 $\pm$ 0.08&	7.53 $\pm$ 0.05 \\
& & & & & \\
{\bf Al I}		&16718.957	&		&		&6.11	&6.12	\\
				&16750.564	&		&		&6.02	&6.01	\\
				&16763.360	&		&		&6.26	&6.37	\\

$\langle$A(Al)$\rangle$ $\pm$ $\sigma$ & & & & 6.13 $\pm$ 0.10&	6.17 $\pm$ 0.15 \\
& & & & & \\
{\bf Si I}		&15888.410	&5.082	&-0.012	&7.31	&7.46	\\
				&15960.063	&5.984	& 0.017	&7.53	&...	\\
				&16094.787	&5.964	&-0.258	&...	&7.70	\\
				&16680.770	&5.984	&-0.190	&7.42	&7.68	\\
                
$\langle$A(Si)$\rangle$ $\pm$ $\sigma$ & & & & 7.42 $\pm$ 0.09&	7.61 $\pm$ 0.11 \\
& & & & & \\
{\bf K I}		&15163.067	&2.670	&0.555	&4.89	&4.82	\\
				&15168.376	&2.670	&0.405	&4.92	&4.83	\\

$\langle$A(K)$\rangle$ $\pm$ $\sigma$ & & & & 4.91 $\pm$ 0.02&	4.83 $\pm$ 0.01 \\
& & & & & \\
{\bf Ca I}		&16136.823	&4.531	&-0.507	&6.32	&6.33	\\
				&16150.763	&4.532	&-0.226	&6.23	&6.31	\\
				&16157.364	&4.554	&-0.169&6.20	&6.25	\\

$\langle$A(Ca)$\rangle$ $\pm$ $\sigma$ & & & & 6.25 $\pm$ 0.05&	6.30 $\pm$ 0.03 \\
& & & & & \\
{\bf Ti I}		&15334.847	&1.887	&-1.040	&4.58	&4.60	\\
				&15543.756	&1.879	&-1.160	&4.70	&7.68	\\
				&15602.842	&2.267	&-1.569 &4.84	&4.78	\\
				&15698.979	&1.887	&-2.110	&4.71	&4.75	\\
				&15715.573	&1.873	&-1.295	&4.55	&4.60	\\
				&16635.161	&2.345	&-1.707	&4.70	&4.86	\\

$\langle$A(Ti)$\rangle$ $\pm$ $\sigma$ & & & & 4.71 $\pm$ 0.08&	4.72 $\pm$ 0.09 \\
& & & & & \\
{\bf V I}		&15924.0	&		&		&3.79	&...	\\

$\langle$A(V)$\rangle$ $\pm$ $\sigma$ & & & & 3.79 $\pm$ 0.01&	... $\pm$ ... \\
& & & & & \\
{\bf Cr I}		&15680.063	&4.697	&0.198	&5.61	&5.60	\\

$\langle$A(Cr)$\rangle$ $\pm$ $\sigma$ & & & & 5.61 $\pm$ 0.01&	5.60 $\pm$ 0.01 \\
& & & & & \\
{\bf Mn I}		&15159.0	&		&		&5.23	&5.31	\\
				&15217.0	&		&		&5.20	&...	\\
				&15262.0	&		&		&5.32	&5.29	\\

$\langle$A(Mn)$\rangle$ $\pm$ $\sigma$ & & & & 5.25 $\pm$ 0.05&	5.30 $\pm$ 0.01 \\

\tablewidth{0pt}	

\enddata

\end{deluxetable}

\begin{deluxetable}{ccccccc}
\tabletypesize{\scriptsize}
%\rotate
\tablecaption{Abundance Sensitivities}
\tablewidth{0pt}
\tablehead{
\colhead{Element} &
\colhead{$\Delta$$T_{\rm eff}$} &
\colhead{$\Delta$log $g$} &
\colhead{$\Delta$[M/H]} &
\colhead{$\Delta$ C/O} &
\colhead{$\Delta$ $\xi$} &
\colhead{$\sigma$}\\
\colhead{} &
\colhead{(+65 K)} &
\colhead{(+0.10 dex)} &
\colhead{(+0.20 dex)} &
\colhead{(+0.15)} &
\colhead{(+0.25 km s$^{-1}$)} &
\colhead{} 
}
\startdata

C  &	+0.01	&	+0.00&	+0.01	&	+0.00	&+0.00	&	0.014\\
O  &	+0.01	&	+0.01&	+0.08	&	+0.01	&-0.02	&	0.084\\
Na &	+0.01	&	+0.00&	-0.07	&	+0.02	&+0.00	&	0.073\\
Mg &	-0.14	&	-0.05&	+0.02	&	-0.02	&+0.00	&	0.151\\
Al &	-0.09	&	-0.05&	+0.02	&	+0.00	&+0.00	&	0.104\\
Si &	-0.14	&	-0.04&	+0.05	&	-0.02	&+0.00	&	0.155\\
K  &	+0.04	&	+0.01&	-0.02	&	+0.01	&+0.00	&	0.047\\
Ca &	-0.01	&	-0.03&	-0.02	&	+0.02	&+0.00	&	0.042\\
Ti &	-0.04	&	-0.03&	-0.07	&	+0.00	&+0.00	&	0.086\\
V  &	+0.01	&	-0.01&	+0.00	&	+0.00	&+0.00	&	0.014\\
Cr &	-0.03	&	-0.02&	-0.02	&	+0.00	&+0.00	&	0.041\\
Mn &	+0.01	&	-0.02&	+0.05	&	+0.00	&+0.00	&	0.054\\
Fe &	-0.08	&	+0.02&	+0.01	&	-0.02	&+0.00	&	0.085\\
$[$C/O$]$  &  +0.00	&	-0.01&	-0.07	&  	-0.01	&+0.02	&	0.074\\
$[$Mg/Si$]$&  +0.00	&	-0.01&	-0.03	&  	+0.00	&+0.00	&	0.031\\

\tablewidth{0pt}	

\enddata

\end{deluxetable}

\begin{deluxetable}{ccccccccc}
\tabletypesize{\scriptsize}
%\rotate
\tablecaption{Mean Abundances and Uncertainties}
\tablewidth{0pt}
\tablehead{
\colhead{Element} &
\colhead{Kepler-138} &
\colhead{$\sigma$}&
\colhead{Kepler-186} &
\colhead{$\sigma$}\\
}
\startdata

$[$C/H$]$  & -0.15 &	0.024	&	-0.08	&0.062	\\
$[$O/H$]$  & -0.16 &	0.086	&	-0.08	&0.103	\\
$[$Na/H$]$ & -0.07 &	0.076	&	-0.01	&0.094	\\
$[$Mg/H$]$ & -0.10 &	0.152	&	+0.00	&0.162	\\
$[$Al/H$]$ & -0.24 &	0.106	&	-0.20	&0.120	\\
$[$Si/H$]$ & -0.09 &	0.156	&	+0.10	&0.166	\\
$[$K/H$]$  & -0.17 &	0.051	&	-0.25	&0.076	\\
$[$Ca/H$]$ & -0.06 &	0.047	&	-0.01	&0.073	\\
$[$Ti/H$]$ & -0.19 &	0.089	&	-0.16	&0.105	\\
$[$V/H$]$  & -0.21 &	0.024	&	 ... 	&...	\\%0.051
$[$Cr/H$]$ & -0.03 &	0.045	&	-0.04	&0.073	\\
$[$Mn/H$]$ & -0.14 &	0.058	& 	-0.09	&0.081	\\
$[$Fe/H$]$ & -0.09 &	0.087	& 	-0.08	&0.104	\\
$[$C/O$]$  &  0.01 &	0.077	&  +0.00	&0.095	\\
$[$Mg/Si$]$& -0.01 &	0.037	&  -0.10	&0.068	\\

\tablewidth{0pt}	

\enddata

\end{deluxetable}

\end{document}